# Exponentially Enhanced non-Hermitian Cooling

Haowei Xu[1], Uroš Delić[2,3], Guoqing Wang[1,3], Changhao Li[1,3],

Paola Cappellaro[1,3,4,†], and Ju Li[1,5,‡]

[1] Department of Nuclear Science and Engineering, Massachusetts Institute of Technology, Cambridge, Massachusetts 02139, USA

[2] University of Vienna, Faculty of Physics, Vienna Center for Quantum Science and Technology, A-1090 Vienna, Austria

[3] Research Laboratory of Electronics, Massachusetts Institute of Technology, Cambridge, MA 02139, USA

[4] Department of Physics, Massachusetts Institute of Technology, Cambridge, MA 02139, USA

[5] Department of Materials Science and Engineering, Massachusetts Institute of Technology, Cambridge, Massachusetts 02139, USA

* Corresponding authors: † pcappell@mit.edu,   ‡ liju@mit.edu

**Abstract**

Certain non-Hermitian systems exhibit the skin effect, whereby the wavefunctions become exponentially localized at one edge of the system. Such exponential amplification of wavefunction has received significant attention due to its potential applications in e.g., classical and quantum sensing. However, the opposite edge of the system, featured by the exponentially *suppressed* wavefunctions, remains largely unexplored. Leveraging this phenomenon, we introduce a non-Hermitian cooling mechanism, which is fundamentally distinct from traditional refrigeration or laser cooling techniques. Notably, non-Hermiticity will not amplify thermal excitations, but rather redistribute them. Hence, thermal excitations can be cooled down at one edge of the system, and the cooling effect can be *exponentially* enhanced by the number of auxiliary modes, albeit with a lower bound that depends on the dissipative interaction with the environment. Non-Hermitian cooling does not rely on intricate properties such as exceptional points or non-trivial topology, and it can apply to a wide range of excitations, such as photons, phonons, magnons, etc.

**Introduction**. Non-Hermiticity arises naturally in open quantum systems. Prototypical examples are loss and gain, which were often considered nuisances in standard Hermitian quantum mechanics. Recently it was realized that a careful balance between gain and loss can lead to the emergence of intriguing phenomena such as parity-time symmetry breaking and exceptional points [1–5]. Other non-Hermitian phenomena, such as non-Hermitian topology [6–11], which can be defined solely based on eigenvalues instead of eigenvectors, have generated considerable attention as well.

A particularly interesting property of certain non-Hermitian systems is the non-Hermitian skin effect (NHSE) [12–19], whereby the wavefunctions are exponentially localized at one boundary of the system. Various novel applications have been proposed based on the NHSE, such as directional amplification of signals [20–23] and (potentially exponentially) enhanced sensing [24–26]. However, the majority of studies in non-Hermitian physics focus on non-Hermiticity-induced *amplification* [1,2,25–28,3,4,15,20–24]. It is worth noting that the exponential amplification of wavefunctions on one edge implies the exponential *suppression* of wavefunctions on the opposite edge. This property, however, remains largely unexplored. In this work, we focus on applications of this effect in non-Hermitian systems, in particular toward cooling thermal excitations.

Cooling down thermal excitations is an essential step for numerous applications that span nearly all scientific and technological domains. However, traditional refrigeration often requires bulky devices, and it can be rather difficult to cool down below certain limits, such as several milli-Kelvin in dilution refrigerators [29]. Meanwhile, laser cooling relies on relatively weak nonlinear optical processes [30–33], necessitating strong pumping lasers that can potentially result in side effects. A critical observation is that in both traditional refrigeration and laser cooling, cooling is achieved by transferring thermal excitations to auxiliary modes (thermal bath) that have *smaller* occupation numbers (Figure 1a, 1b). This raises the question: is this a necessary condition for cooling?

In this work, we propose a non-Hermitian cooling mechanism, which is related to non-reciprocal refrigeration [34,35], but is fundamentally different from traditional refrigeration or laser cooling. The non-Hermitian cooling stems from the *directional* transport of thermal excitations and can be achieved even if the auxiliary modes have the same occupation number (Figure 1c). We will demonstrate that directional transport does not significantly amplify but only redistributes thermal excitations in non-Hermitian systems. This guarantees that the localization of thermal excitation and heating on one edge (i.e., NHSE) will at the same time result in cooling on the opposite edge. By employing two different theoretical approaches, we reveal that the cooling effect can be exponentially enhanced by the number of auxiliary non-Hermitian modes, although a lower bound emerges when the dissipative interaction with the environment is included.

The non-Hermitian cooling is inherently versatile. In principle, it applies to various types of excitations including photons, phonons, magnons, etc. The non-Hermitian cooling does not require intricate properties such as parity-time symmetry [1,36–38], exceptional points [1–5], or non-trivial topologies [15,25,26]. The only requirement is the non-reciprocal hopping and directional transport of the excitations, corresponding to the emergence of the NHSE. Hence, we expect non-Hermitian cooling to be widely applicable.

In the following, we will first explain the distinctiveness of non-Hermitian cooling by comparing it with traditional refrigeration and laser cooling methods. Then, starting from two-mode systems, we will demonstrate the performance of non-Hermitian cooling, including the exponential enhancement in multi-mode systems, as well as the lower bound of the cooling effect. Finally, we will discuss some issues relevant to the practical realization of non-Hermitian cooling.

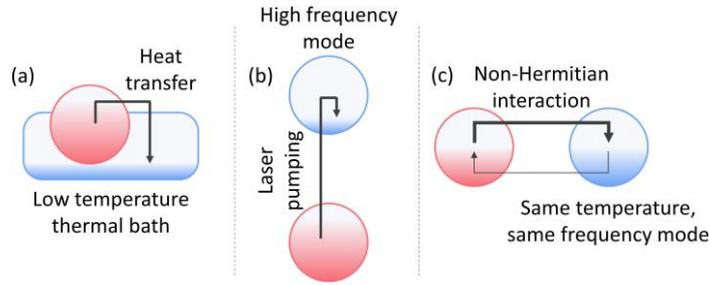

**Figure 1.** Illustration of the mechanism of (a) traditional refrigeration, (b) laser cooling, and (c) non-Hermitian cooling. Red (blue) circles denote the principal (auxiliary) mode, and the color filling denotes thermal occupation. The arrows indicate the transfer of thermal excitation, with thicker arrows corresponding to faster transfer.

**Traditional refrigeration, laser cooling, and non-Hermitian cooling.** For clarity, we will use principal and auxiliary to denote the mode to be cooled down and the modes that facilitate the cooling process, respectively. For traditional refrigeration, the key step is to attach the principal mode to the auxiliary modes (thermal bath) with the same intrinsic frequency but lower temperature, which have smaller occupation numbers. Then, the thermal excitations would naturally flow into the thermal bath via heat transfer, leading to the cooling effect (Figure 1a). Clearly, the lowest achievable temperature is the temperature of the thermal bath. While diverse schemes have been devised to create thermal baths with low temperatures, it can be extremely demanding and costly for traditional refrigeration to go beyond a certain limit, e.g., several milli-Kelvin in dilution refrigerators [29].

Laser cooling uses a different mechanism – a low-frequency principal mode is coupled to an auxiliary mode with a higher (usually optical) frequency, whose effective thermal occupation is much smaller even at elevated temperatures (Figure 1b). It is worth noting that some other processes, such as dynamic nuclear polarization [39], employ a similar mechanism. For laser cooling, the thermal excitations in the principal mode are pumped to the auxiliary mode by an external laser, which compensates for the energy difference. While laser cooling is remarkably successful, as seen in the achievement of picokelvin temperatures for ultracold atoms [40], it is ultimately limited by the laser power and the dissipation of the principal and auxiliary modes. Specifically, the transfer of thermal excitations via laser pumping relies on intrinsically weak and usually nonlinear optical processes. Hence, strong pumping lasers are required, which can cause heating and damage to the surrounding, especially in solid-state systems. Moreover, the auxiliary high-frequency modes usually have high dissipation rates as well, which also limits the effectiveness of laser cooling.

One can see that both traditional refrigeration and laser cooling result from transferring thermal excitations to auxiliary modes with smaller occupation numbers. Indeed, for bosonic modes, the transition (energy transfer) rate from the principal to the auxiliary mode is $gn_1(1 + n_2)$, while that in the reverse direction is $gn_2(1 + n_1)$, so the net flow is $g(n_1 - n_2)$. Here $n_1$ ($n_2$) is the occupation number of the principal (auxiliary) mode, while $g$ is the reciprocal coupling strength. Clearly, cooling of the principal mode requires $n_2 < n_1$. Similar analyses hold for fermionic modes as well.

Nevertheless, this is not a necessary condition for cooling in a non-Hermitian system. Even if the principal and auxiliary modes have the same occupation number, cooling could still be achieved if (1) the transfer of excitations is directional, so that the excitations preferably jump from the principal to the auxiliary modes (Figure 1c); (2) the directional transport does not significantly amplify total thermal excitations in the system. In the following, we will demonstrate that both conditions can be satisfied in certain non-Hermitian systems.

**Non-Hermitian cooling in two-mode systems.** To better understand the non-Hermitian cooling effect, let us first examine a two-mode system, where the principal mode-1 and auxiliary mode-2 have the same intrinsic frequency $\omega_0$ (inset of Figure 2b). We will assume the two modes to be bosons, such as photons, phonons, or magnons, but the discussions below can be adapted to fermionic modes as well. We will show that in a non-Hermitian system, the transition rate from modes 1 to 2 can be different from that in the reverse direction, resulting in an unbalanced steady-state occupation (SSO) and an effective cooling effect. We consider the following Hamiltonian of the two-mode system in the rotating frame of $\omega_0$

$$\mathcal{H} = t_{12} a_1 a_2^\dagger + t_{21} a_1^\dagger a_2. \tag{1}$$

Here $a_i$ ($a_i^\dagger$) is the annihilation (creation) operator of mode-$i$, while $t_{12} = te^A$ and $t_{21} = te^{-A}$ are the inter-mode coupling strength. The interaction is non-Hermitian when $A$ has a non-zero real part. For definiteness and without loss of generality, we will take $A > 0$ as a real number unless explicitly stated. The time evolution of the density matrix $\rho$ can be described by the quantum master equation [41–43], $\frac{\partial \rho}{\partial t} = -i(\mathcal{H}\rho - \rho \mathcal{H}^\dagger) + i\text{Tr}\{\rho(\mathcal{H} - \mathcal{H}^\dagger)\}\rho + \sum_i \mathcal{L}(o_i)\rho$, which is adapted for a non-Hermitian Hamiltonian $\mathcal{H} \neq \mathcal{H}^\dagger$ (see Section 1 of Ref. [44] for detailed discussions, which also includes Refs. [8,14,21,41,45–47]). The dissipative interactions with the environment (thermal bath) are described by the Lindblad operators $\mathcal{L}(o_i)\rho \equiv o_i \rho o_i^\dagger - \frac{1}{2}(o_i^\dagger o_i \rho - \rho o_i^\dagger o_i)$ for operator $o_i$.

One can readily observe that when the dissipative interactions with the thermal bath are ignored, the "Rabi" oscillation between modes 1 and 2 is still periodic, but will deviate from a normal sinusoidal function (Figure 2a). Compared with the $1 \to 2$ oscillation, the reverse $2 \to 1$ oscillation takes a longer time $\tau_{21} \sim \frac{1}{t_{21}}$, resulting in an unbalance between the two modes, and the excitations would preferably reside in mode 2.

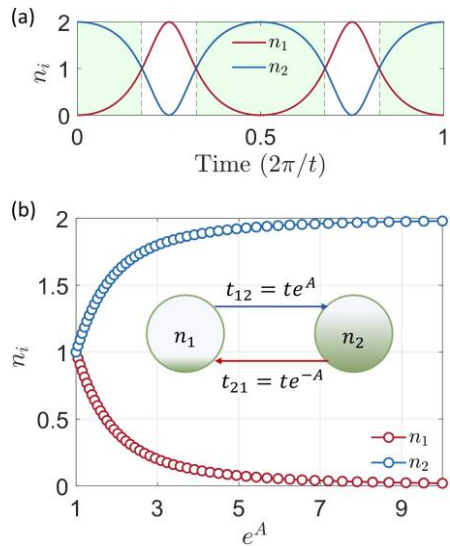

**Figure 2**. (a) Unbalanced "Rabi" oscillation in the non-Hermitian two-mode system with $e^A = 2$. Green and white shaded regions correspond to $n_1 < n_2$ and $n_1 > n_2$, respectively. (b) Steady-state

occupation as a function of $e^A$ with $\kappa_1 = \kappa_2 = 0.01\,t$. Inset of (b): Illustration of a non-Hermitian two-mode system

Considering the dissipative interactions with the thermal bath and assuming the occupation numbers to be small, one has (Section 1 of Ref. [44])

$$\frac{\partial n_1}{\partial t} \approx i\big[t_{12}\langle a_1 a_2^\dagger\rangle - t_{21}\langle a_1^\dagger a_2\rangle\big] - \kappa_1[n_1 - n_{\text{th}}],$$
$$\frac{\partial \langle a_1^\dagger a_2\rangle}{\partial t} \approx i\big[t_{21}^*\langle a_2^\dagger a_2\rangle - t_{12}\langle a_1^\dagger a_1\rangle\big] - \frac{\kappa_1+\kappa_2}{2}\langle a_1^\dagger a_2\rangle. \quad (2)$$

Here $\kappa_i$ is the dissipation rate of mode-$i$, and $n_{\text{th}}$ is the occupation number of the thermal bath, which is assumed to be the same for the two modes. Note that $n_{\text{th}}$ is also the equilibrium occupation number of the two modes if the non-Hermiticity is absent (i.e., $t_{12} = t_{21}$). Meanwhile, $\langle o \rangle \equiv \text{Tr}\{\rho o\}$ indicates the thermal average of operator $o$, and $n_1 \equiv \langle a_1^\dagger a_1\rangle$ is the occupation of mode-1. Similar equations hold when modes 1 and 2 are exchanged. In the steady state, one has

$$g_{12} n_1 + \kappa_1 n_1 \approx g_{21} n_2 + \kappa_1 n_{\text{th}},$$
$$g_{21} n_2 + \kappa_2 n_2 \approx g_{12} n_1 + \kappa_2 n_{\text{th}}, \quad (3)$$

where $g_{ij} = \frac{2(|t_{ij}|^2 + t_{ij} t_{ji})}{\kappa_i+\kappa_j}$ is the non-Hermitian transition rate from mode $i$ to $j$. Note that one recovers $g_{ij} = \frac{4t^2}{\kappa_i+\kappa_j}$ for $A = 0$, which is a well-known result obtained from Fermi's golden rule.

The SSO $n_1$ and $n_2$ as a function of $e^A$ are shown in Figure 2b, where we assume $\kappa_1 = \kappa_2 = 0.01t$ and $n_{\text{th}} = 1$. In a Hermitian system with $A = 0$, one has $n_1 = n_2 = n_{\text{th}}$, as expected. In contrast, one has $n_1 < n_{\text{th}}$ for $A > 0$, which is the non-Hermitian cooling effect. Moreover, $n_1$ gradually decreases to zero as $e^A$ increases. Nonetheless, for finite values of $e^A$, the cooling effect is limited. For example, one has $n_1 \approx 0.4\, n_{\text{th}}$ when $e^A = 2$. Note that this cooling effect is closely related to non-reciprocal refrigeration [34,35].

**Exponential non-Hermitian cooling in multi-mode systems.** The cooling effect described in the previous section can be exponentially enhanced if more auxiliary modes are added. Ideally, one should be able to achieve $n_1 \propto \left(\frac{g_{12}}{g_{21}}\right)^{-N}$ in a $N$-mode non-Hermitian chain (Figure 3a). The Hamiltonian of the chain in the rotating frame is

$$\mathcal{H} = \sum_{i=1}^{N-1} t_{i,i+1} a_i a_{i+1}^\dagger + t_{i+1,i} a_i^\dagger a_{i+1} \qquad (4)$$

with $t_{i,i+1} = te^A$ and $t_{i+1,i} = te^{-A}$. This is the renowned non-Hermitian Hatano-Nelson model [12,18]. By diagonalizing Eq. (4), one has $\mathcal{H}|\psi_\alpha\rangle = \epsilon_\alpha |\psi_\alpha\rangle$, where the $\alpha$-th eigenstate has energy $\epsilon_\alpha$ and right eigenvector $|\psi_\alpha\rangle$. Under the current setting, $\mathcal{H}$ can be mapped to a Hermitian Hamiltonian by a similarity transformation [8,48]. Consequently, $\epsilon_\alpha$ are all real numbers, while $|\psi_\alpha\rangle$ are exponentially localized on the right boundary, i.e., $\psi_\alpha^i \approx e^A \psi_\alpha^{i-1}$, which is the NHSE [8,18]. In many scenarios, the coupling strengths $te^{\pm A}$ are much smaller than other energy scales, such as the intrinsic mode frequency $\omega_0$ and the temperature $T$. Hence, if one arbitrarily applies standard thermodynamics to the Hatano-Nelson model [49], then all the eigenstates $|\psi_\alpha\rangle$ should have almost the same probability to be occupied, and the SSO of the $i$-th mode is

$$n_i^{\text{HN}} = n_{\text{th}} \sum_\alpha |\psi_\alpha^i|^2. \qquad (5)$$

Here the superscript HN indicates that the occupations are obtained directly from the wavefunctions of the Hatano-Nelson model. One can easily verify that $n_i = n_{\text{th}}$ if the Hamiltonian is Hermitian ($A = 0$), as expected. In contrast, with $A > 0$, the NHSE implies that the SSO on the left edge is exponentially suppressed with $n_i^{\text{HN}} \approx e^{-2A} n_{i+1}^{\text{HN}}$ and $n_1^{\text{HN}} \sim n_{\text{th}} e^{-2NA}$, corresponding to the exponential cooling effect.

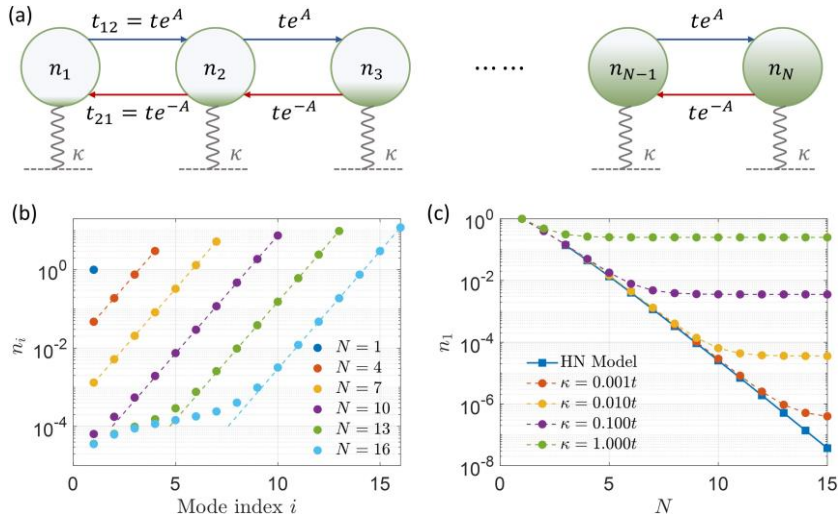

**Figure 3.** Illustration of a non-Hermitian $N$-mode chain. The color filling denotes SSO. (b) SSO of each mode in the chain with varying length $N$. The dashed line indicates the exponential scaling of $n_i \propto \left(\frac{g_{12}}{g_{21}}\right)^{N-i}$. (c) SSO of the leftmost mode-1 as a function of $N$ with varying $\kappa$. The solid blue line comes from Eq. (5), which coincides with the case of $\kappa \to 0$. We set $e^A = 2$ in (b, c) and $\kappa = 0.01\, t$ in (b).

One may wonder whether the exponential cooling effect described above can survive if the dissipative interactions with the environment are incorporated. Fortunately, the answer is yes, although a lower bound on $n_1$ would arise with given $e^A$ and $\kappa$. Generalizing the steady-state equations Eq. (3) to the $N$-mode system, one has

$$g_{i,i+1}n_i + g_{i,i-1}n_i + \kappa n_i = g_{i+1,i}n_{i+1} + g_{i-1,i}n_{i-1} + \kappa n_{\text{th}}, \quad i = 1,2,\ldots,N \quad (6)$$

where for simplicity, we assume the thermal dissipation rate ($\kappa$) and the thermal bath occupation ($n_{\text{th}}$) to be the same for all modes. Meanwhile, we set $g_{i,i+1} = g_{12}$ and $g_{i+1,i} = g_{21}$, and the open boundary condition is implicitly incorporated by setting $g_{01} = g_{10} = g_{N,N+1} = g_{N+1,N} = 0$.

An intriguing and important observation from Eq. (6) is that the sum of SSO on all modes remains a constant, i.e., $\sum_i n_i \equiv N n_{\text{th}}$, regardless of the values of $g_{ij}$. This implies that an increase in the SSO of some modes must be accompanied by a decrease in other modes. While the thermal excitations can internally redistribute in different modes, the $N$-mode system as a whole must exhibit $N n_{\text{th}}$ excitations to the environment. This should be compared with the cases whereby an input signal is amplified by the non-Hermitian chain [20–22], which can be used to improve classical and/or quantum sensing [25,26]. Here, the non-Hermitian chain does not amplify but only redistributes the thermal excitations. This crucial property guarantees that the NHSE will lead to exponential non-Hermitian cooling.

We numerically solve the steady-state equations Eq. (6) to obtain the SSO in a non-Hermitian chain (Figure 3b, 3c). $e^A$ is set to 2 unless explicitly stated. With a fixed and finite $\kappa$, the SSO of the leftmost mode $n_1$ first decreases exponentially with $N$, and then becomes a constant $\lim_{N \to \infty} n_1(\kappa, N) \approx \frac{\kappa^2 n_{\text{th}}}{\kappa^2 + t_{12}^2 - t_{21}^2}$ (Section 2 of Ref. [44]). While obtained using a different approach, $n_1^{\text{HN}}$ in Eq. (5) coincides with the case of $\kappa \to 0$, i.e., $n_1^{\text{HN}} \approx \lim_{\kappa \to 0} n_1(\kappa, N)$. This agreement suggests that the theoretical results presented here should be robust. The non-reciprocal cooling studied in Refs. [34,35] correspond to the special case of $N = 2$ in the current work. We would like to emphasize that using a multi-mode system ($N > 2$) can further reduce the lowest achievable occupation number $n_1$ of the principal mode.

**Cooling down a reciprocally coupled mode.** In some situations, establishing a non-Hermitian interaction can be challenging, and/or can lead to unwanted side effects for the principal mode, such as extra dissipations [45]. In this case, one can attach the principal mode (denoted by 0 here) to mode-1 in the non-Hermitian chain described above. The interaction between modes 0 and 1, $\mathcal{H}_0 = t_0(a_0 a_1^\dagger + a_0^\dagger a_1)$, can be Hermitian (Figure 4a). The cooling effect of mode-0 comes from the fact of $n_1 \ll n_{\text{th}}$, as the flow of thermal excitation will be dominantly in the $0 \to 1$, rather than the $1 \to 0$ direction, even if the transition between modes 0 and 1 is reciprocal. Using the steady-state equations Eq. (6), one can obtain the SSO (Figure 4b)

$$n_0 \approx \frac{g_0 n_1 + \kappa_0 n_{\text{th}}}{g_0 + \kappa_0}, \tag{7}$$

where $\kappa_0$ is the dissipation rate of mode 0, and $g_0 = \frac{4|t_0|^2}{\kappa + \kappa_0}$ is the reciprocal transition rate between mode 0 and 1. One can see that a cooling effect ($n_0 < n_{\text{th}}$) can be realized whenever $n_1 < n_{\text{th}}$.

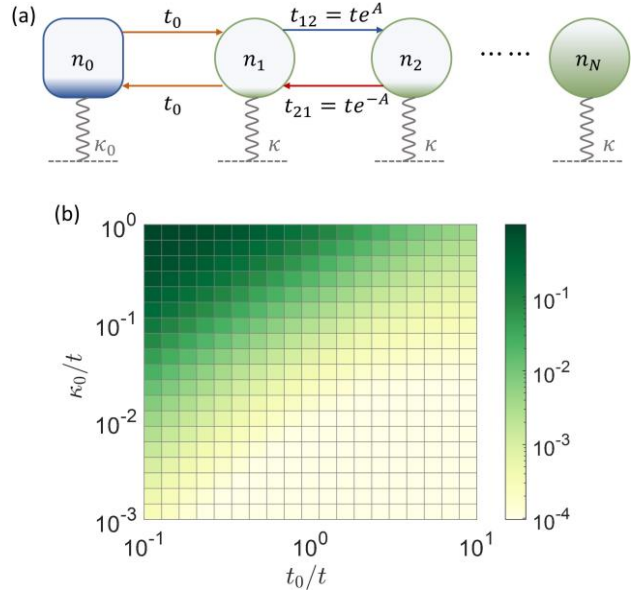

**Figure 4.** (a) A Hermitian principal mode-0 is attached to the non-Hermitian chain. (b) SSO of mode-0 as a function of $\kappa_0$ and $t_0$. The non-Hermitian chain has $N = 15$ modes with $e^A = 2$ and $\kappa = 0.01\, t$.

**Discussions**. The non-Hermitian cooling efficiency depends on two parameters, namely $e^A$ and $\frac{\kappa}{t}$. In theory, one can achieve $e^A \to \infty$ by judiciously designing the interaction between the two modes using e.g., reservoir engineering [45] or parametric driving [21,46]. In practice, $e^A \gtrsim 3$ has been realized in e.g., Josephson junctions [20], optomechanical circuits [50], and optically levitated nanoparticles [51]. Hence, we believe $e^A = 2$ used in this work is feasible. The non-Hermitian interaction between multiple modes has been demonstrated as well [52]. Still, additional experimental efforts are required to further increase $e^A$ and $N$, so that the potential of the non-Hermitian cooling effect can be fully exploited. A concern can arise from the possible detrimental influence of thermal fluctuations in the externals modes or driving fields that are used to induce the non-Hermitian interactions. Fortunately, our analyses show that the influence of such thermal fluctuations should be negligible in generating the non-Hermitian dynamics, at least in the first order approximation (Section 4 of Ref. [44]). Moreover, in practice the non-Hermitian interactions are often driven by optical photons [51,53], whose thermal fluctuations are extremely small because of their high intrinsic frequencies. Furthermore, strong couplings ($\frac{\kappa}{t} \ll 1$) can be realized between microwave resonators [54], mechanical resonators [55], magnons [56,57], etc. It should be emphasized that the coupling $t$ here can be linear interactions without external pumping, such as the Zeeman interaction between magnons and microwave resonators. In contrast, for laser cooling the interactions between the principal and auxiliary modes are relatively weak (nonlinear) optical processes under external laser pumping, which leads to some limitations discussed before.

The presented non-Hermitian cooling applies to various excitations, including photons [15,58], phonons [17,50,59,60], Josephson circuits [20,21], magnons [42,61], etc., whereby non-Hermitian interactions have been realized. Moreover, non-Hermitian cooling exists whenever the wavefunction density is suppressed in certain local regions (not necessarily on the edge) of the non-Hermitian system. It does not require sophisticated properties such as exceptional points [1–5], non-trivial topology [15,26], or mixing between different quadratures of the electromagnetic fields [25]. As an example, we investigated the non-Hermitian Su–Schrieffer–Heeger model (Section 3 of Ref. [44]), which can undergo topological phase transitions under certain conditions [8,18,62]. We find that non-Hermitian cooling exists and is almost identical in both topologically trivial and non-trivial phases.

In summary, we proposed a non-Hermitian cooling mechanism that is fundamentally different from traditional refrigeration and laser cooling techniques. The non-Hermitian cooling is generically applicable to various physical systems and can be exponentially enhanced by the number of auxiliary modes. We anticipate the non-Hermitian cooling paradigm to find applications in various domains, such as quantum information science, where cooling towards the ground state is desirable.

# Acknowledgment

We acknowledge support by the Gordon and Betty Moore Foundation's EPiQS Initiative, Grant GBMF11945.